%% file: TopmetalII-.tex
\documentclass[preprint,5p,number,sort&compress]{elsarticle}
\usepackage{dcolumn}
\usepackage{bm}
\usepackage{upgreek}
\usepackage{graphicx}
\usepackage{hyperref}
\usepackage[load-configurations=abbreviations,detect-all=true]{siunitx}
\usepackage[version=3]{mhchem}
\usepackage[mathlines]{lineno}
\usepackage[usenames,dvipsnames,svgnames,table]{xcolor} 
\usepackage[normalem]{ulem} 
\usepackage{transparent}

\input{defs.tex}

\journal{Nuclear Instruments and Methods in Physics Research A}

\begin{document}

\begin{frontmatter}

  \title{A Low-Noise CMOS Pixel Direct Charge Sensor, \TMIIm}

\author[ccnu]{Mangmang An}
\author[ccnu]{Chufeng Chen}
\author[ccnu]{Chaosong Gao}
\author[lbnl]{Mikyung Han}
\author[ccnu]{Rong Ji}
\author[ccnu]{Xiaoting Li}
\author[lbnl]{Yuan Mei\corref{cor0}}\ead{ymei@lbl.gov}
\author[ioa]{Quan Sun}
\author[ccnu]{Xiangming Sun\corref{cor0}}\ead{xmsun@phy.ccnu.edu.cn}
\author[ccnu]{Kai Wang}
\author[ccnu]{Le Xiao}
\author[ccnu]{Ping Yang}
\author[ccnu]{Wei Zhou}

\address[ccnu]{Central China Normal University, Wuhan, Hubei 430079, China}
\address[lbnl]{Nuclear Science Division, Lawrence Berkeley National Laboratory, Berkeley,
  California 94720, USA}
\address[ioa]{Institute of Acoustics, Chinese Academy of Sciences, Beijing 100190, China}

\cortext[cor0]{Corresponding author}

\begin{abstract}

  We report the design and characterization of a CMOS pixel direct charge sensor, \TMIIm,
  fabricated in a standard \SI{0.35}{\micro\meter} CMOS Integrated Circuit process.  The sensor
  utilizes exposed metal patches on top of each pixel to directly collect charge.  Each pixel
  contains a low-noise charge-sensitive preamplifier to establish the analog signal and a
  discriminator with tunable threshold to generate hits.  The analog signal from each pixel is
  accessible through time-shared multiplexing over the entire array.  Hits are read out digitally
  through a column-based priority logic structure.  Tests show that the sensor achieved a
  $<\SI{15}{e^-}$ analog noise and a \SI{200}{e^-} minimum threshold for digital readout per
  pixel.  The sensor is capable of detecting both electrons and ions drifting in gas.  These
  characteristics enable its use as the charge readout device in future Time Projection Chambers
  without gaseous gain mechanism, which has unique advantages in low background and low
  rate-density experiments.

\end{abstract}

\end{frontmatter}

\section{Introduction}\label{sec:introduction}

Over the years a number of highly pixelated CMOS sensors and readout Integrated Circuit (IC)
chips have been developed and deployed successfully in nuclear and particle physics experiments.
Notable examples are CMOS readout pixel ICs such as
Medipix/Timepix\cite{Ballabriga2011S15,Llopart2007485} and FE-I3/FE-I4 \cite{FE-I4}, and
Monolithic Active Pixel Sensors (MAPS) such as the one described in \cite{HuGuo2010480}.  Readout
pixel ICs are designed mainly to be coupled to external solid state detectors through processes
such as flip-chip bump bonding.  Charges are generated in the external detector due to the
passing of ionizing particles, then collected and measured by the readout IC.  MAPS ICs allow
charges to be generated inside of the silicon IC (pixel) itself and integrate the readout
circuitry in the same chip.

The high pixel density and high integration of circuitry nature of pixel ICs made them appealing
to be the charge readout device of choice in Micro-Pattern Gaseous Detectors
(MPGDs)\cite{Titov2013}, replacing conventional wire readout schemes\cite{Charpak1978} and
patterned Printed Circuit Board (PCB) readout schemes\cite{Nagayoshi200420}.  These MPGDs are
usually used for charge and timing measurements in Time Projection Chambers (TPCs)\cite{TPC}
achieving improved spatial resolution and noise performance.  For instance, the D$^3$ experiment
\cite{d3} uses FE-I3/FE-I4 sensors developed for the ATLAS \cite{atlas} experiment, behind a
Gaseous Electron Multiplication (GEM) stage, to detect charge tracks resulting from potential
dark matter interactions.  Some similar efforts using the Timepix sensor behind GEMs were
presented in \cite{Campbell2005295,GEM-TPC-PXL}.  A more integrated approach, fabricating a
micro-pattern gaseous gain structure directly on top of a Medipix/Timepix chip by means of wafer
post-processing, was reported in \cite{vanderGraaf20095} (InGrid).

In these applications of pixel ICs in gaseous detectors, ICs were usually designed for a
different purpose and were later converted to perform direct charge readout resulting in
characteristic mismatches between the IC performance and application requirements.  Very few
direct charge collection ICs were designed specifically for micro-pattern gaseous pixel
detectors.  A notable exception is an IC described in \cite{Bellazzini2004477} that is dedicated
to GEM readout and X-ray polarimetry applications\cite{Bellazzini2006425}.

Regardless of which IC is employed for charge readout, an electron gas-avalanche gain stage is
involved in all the above mentioned pixel readout systems.  These systems are generally geared
towards detectors for high event rate and high electron drift speed.  An electron gas-avalanche
gain is necessary to amplify the number of electrons to be well above the noise of ICs.  Also, a
pulse shaper is normally built in-chip to increase the rate capabilities.  On the other hand,
there is a class of measurements that disfavor the use of gas-avalanche gain while demand similar
spatial and timing resolution.  They are usually low rate-density (event rate per volume) and low
background experiments.  Examples are alpha particle counting\cite{Gordon2009}, neutrinoless
double-beta decay (\znbb) searches by drifting ions in a high pressure
TPC\cite{Chinowsky2007,Nygren2013}, upgrades to ion chambers for beam measurements, and advanced
electron-track gamma-ray imaging\cite{Haefner2014}.  Gas-avalanche gain is disfavored due to its
large gain fluctuation and sparking resulting in poor energy resolution and stability issues
particularly in large-area readout systems.  Also, positive ions and negative ions in high
pressure gas cannot undergo gas-avalanche.  High spatial resolution is required for imaging,
position determination and interaction vertex identification.

We designed a CMOS IC, \TMIIm, that is uniquely suitable for charge measurement in a TPC without
gas-avalanche gain.  It is a direct charge sensor with \SI{83}{\micro m} pitch between pixels
fabricated in a standard \SI{0.35}{\micro m} CMOS process without post-processing.  A metal patch
is placed on the top of each pixel (\TM) in a $72\times72$ pixel array for charge collection.
Each pixel contains a low-noise charge-sensitive preamplifier (CSA) to establish the analog
signal and a discriminator with tunable threshold to generate hits.  For its intended
application, event-rate density is expected to be low and charge (both free electron and ion)
drifting speed is expected to be slow; therefore, we tuned the preamplifier to have long signal
retention and eliminated the in-chip pulse shaper while focusing on improving the noise
performance.  The analog signal from each pixel is accessed through time-shared multiplexing over
the entire array.  Hits are read out digitally through a column-based priority logic structure.

In this paper we present the overall design and some initial test results of \TMIIm.  The
implementation of the charge collection electrode and a time-shared analog multiplexing structure
have been validated in our earlier IC development \TMI\cite{TopmetalI2014}.  We will focus on the
analog characteristics from CSA and the behavior of digital readout.  Details on circuitry and
specific application notes will be presented elsewhere.

\section{Sensor Structure and Operation}\label{sec:sensor-structure-operation}

\begin{figure}[htbp]
  \centering
  \begin{minipage}[c]{0.4\linewidth}
    \vspace{7ex}
    \includegraphics[width=\linewidth]{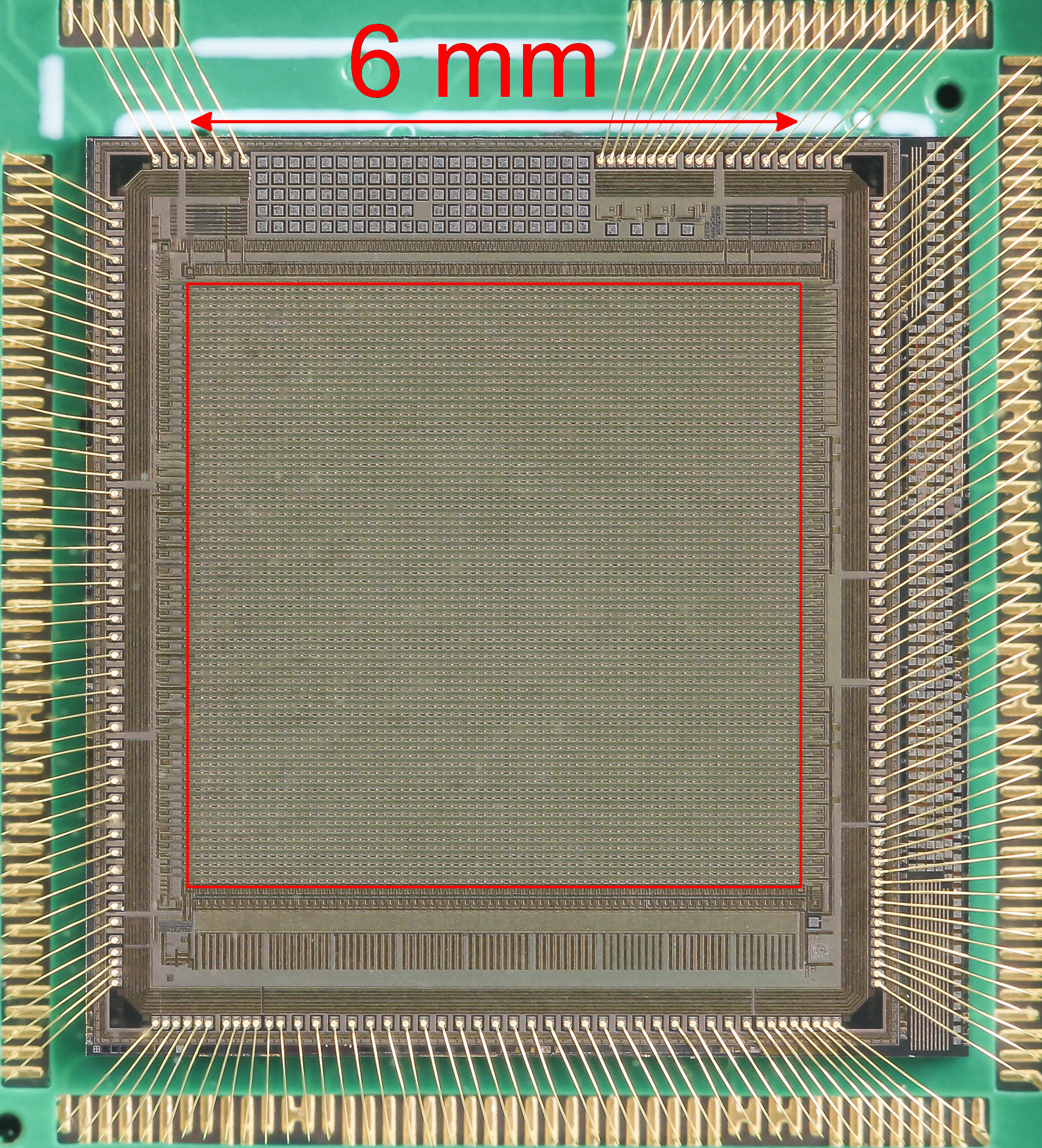}
  \end{minipage}%
  \begin{minipage}[c]{0.6\linewidth}
    \vspace{5ex}
    \includegraphics[width=\linewidth]{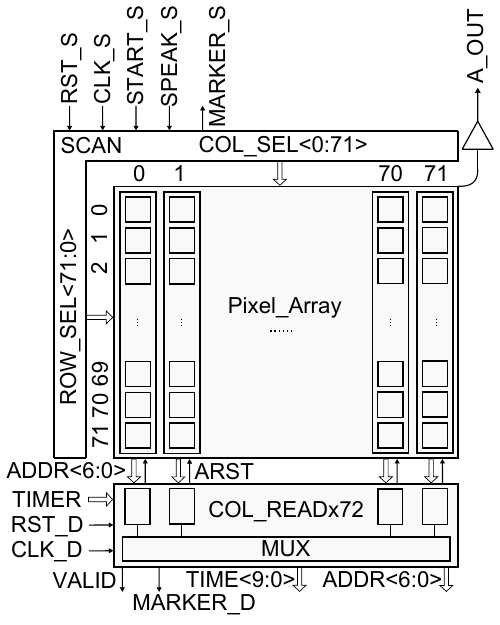}
  \end{minipage}
  \caption{Photograph of a \TMIIm sensor (left) and a schematic view of the overall architecture
    (right).  The sensor chip is placed on a PCB and gold wire-bonded.  The $72\times72$ pixel
    array, with $\sim\SI{83}{\micro m}$ pitch distance between pixels, constitutes an
    approximately $6\times\SI{6}{mm^2}$ charge sensitive area (red box) in the center of the
    sensor.  The analog output from each pixel is fed to a single output buffer via an array-wide
    row/column multiplexing circuitry.  The digital output (hits) is registered at the bottom of
    each column, then accessed by column polling.}
  \label{fig:TMIImPhoto} \label{fig:TMIImOverall}
\end{figure}

A photograph of one wire-bonded \TMIIm sensor is shown in Fig.~\ref{fig:TMIImPhoto} (left).  The
sensor is implemented in a $8\times\SI{9}{mm^2}$ silicon real-estate area.  A schematic view of
the sensor architecture is shown in Fig.~\ref{fig:TMIImOverall} (right).  With
$\sim\SI{83}{\micro m}$ pitch distance between pixels, the $72\times72$ square pixel array makes
up a $6\times\SI{6}{mm^2}$ charge sensitive region.  Readout interface logic and an analog buffer
are placed adjacent to the pixel array.  The sensor is powered by analog supply \sym{AVDD} and
digital supply \sym{DVDD}, which are individually regulated at a nominal voltage of \SI{3.3}{V}.

Each pixel collects charges via its own $25\times\SI{25}{\micro m^2}$ sized metal node (\TM),
then converts them to both analog and digital signals (Fig.~\ref{fig:SinglePixelStruct}).
Separate circuit structures read out both analog and digital signals from every pixel and send
them through dedicated channels to external interfaces.  The pixel structure, analog, and digital
readouts are described in the following subsections.

\subsection{Pixel Structure}

\begin{figure}[htbp]
  \centering
  \includegraphics[width=\linewidth]{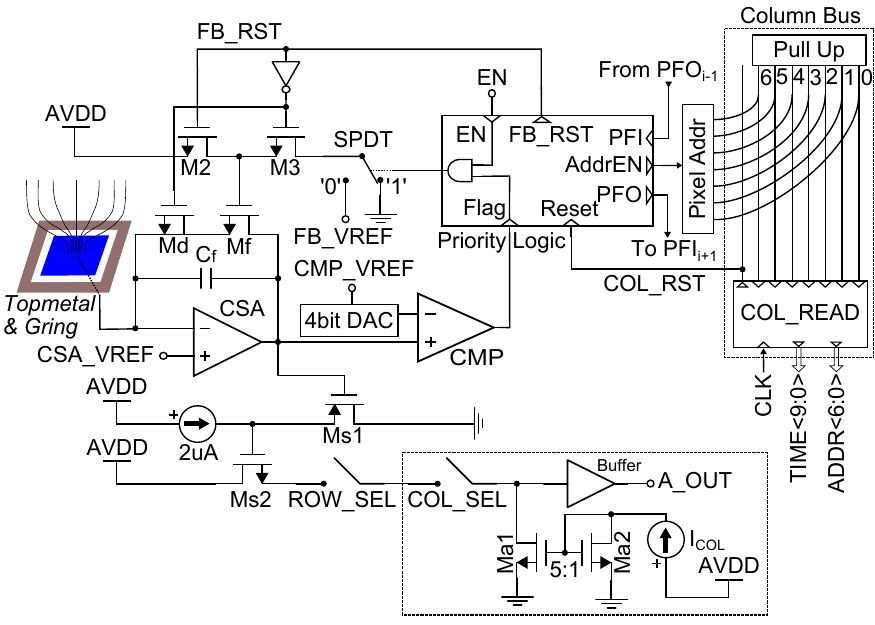}
  \caption{Internal structure of a single pixel.  Charges are collected on the \TM and the CSA
    converts them into an analog voltage signal.  The analog signal is routed through an
    array-wide multiplexer to be accessed externally.  The analog signal is also fed into a
    comparator (CMP) to generate hits, which is then readout digitally via a column-based
    priority logic.  Dashed boxes indicate structures external to the pixel, which are shared
    either at column level or at array level.}
  \label{fig:SinglePixelStruct}
\end{figure}

A schematic view of the internal structure of a single pixel is shown in
Fig.~\ref{fig:SinglePixelStruct}.  The \TM is directly connected to the input of the CSA.  Around
the \TM there is a ``Guard Ring'' (\sym{Gring}), which is a ring electrode in the same topmost
metal layer as the \TM but isolated from it.  \sym{Gring} serves a dual purpose.  Firstly, when
biased to a different electric potential than the \TM, an electric field gradient is formed
around the \TM, which focuses the moving charges hence enhances the charge collection efficiency.
Secondly, the stray capacitance between the \sym{Gring} and the \TM, \Cgt, is a natural test
capacitor.  It allows us to apply pulses on Gring to inject signals into the CSA to measure its
performance.  The parametric extraction from the IC design software as well as an independent
Finite Element Analysis (FEA) show that $\Cgt\approx\CgtVal$.

\begin{figure}[htbp]
  \centering
  \includegraphics[width=0.7\linewidth]{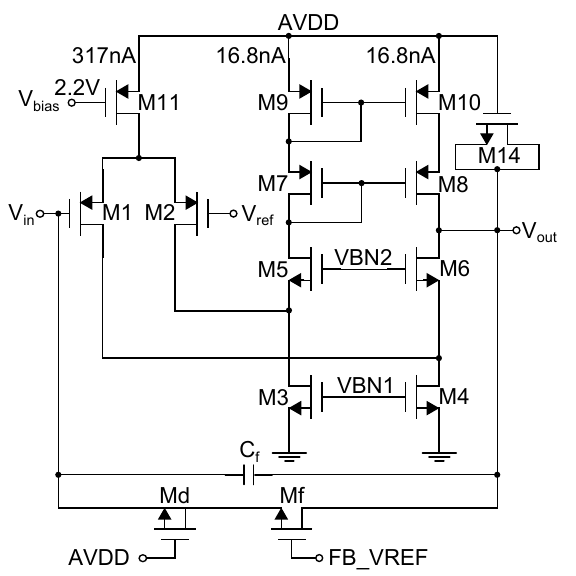}
  \caption{Internal structure of the Charge Sensitive pre-Amplifier (CSA).  Voltages and currents
    marked are typical values during analog-only operations.}
  \label{fig:CSAStruct}
\end{figure}

The CSA is a dual-input folded cascode Operational Amplifier (OpAmp) with capacitive feedback
(Fig.~\ref{fig:CSAStruct}).  The parasitic capacitance of the \TM to ground, plus the gate
capacitance of M1, is about \CinVal, which is presented to the CSA as its input capacitance.  The
reference voltage $V_\text{ref}=\sym{CSA\_VREF}$ sets the baseline voltage of the \TM.  The
feedback capacitor $C_f$ is implemented using the stray capacitance between two metal traces.  We
estimate $C_f\approx\SI{5}{fF}$ from parametric extraction in the IC design software.  During
normal analog-only operations (\sym{EN}=0,\sym{FB\_RST}=0), a fine-adjusted voltage
\sym{FB\_VREF} is applied through M3 in Fig.~\ref{fig:SinglePixelStruct} onto the gate of Mf,
which sets the equivalent feedback resistance $R_f$ ($R_\text{DS}$ of Mf).  The gate-source
voltage of Mf is determined by $\Vgsf=\sym{FB\_VREF}-\sym{CSA\_VREF}$.  We usually set \Vgsf to a
few hundred \si{mV}, which is sub-threshold for Mf and results in a large $R_f$.  Consequently,
the feedback time constant $\tau_f=R_f\cdot C_f$ can be as large as seconds, and it can be
adjusted by varying \Vgsf.  Md is a ``dummy'' transistor that does not affect normal analog
operations.  However during digital readout, we utilize its gate capacitance to counteract the
charge injection from Mf during switching.

We designed the CSA to be particularly low in power consumption ($<\SI{1}{\micro A}$ per CSA) and
intentionally limited its bandwidth to below \SI{1}{MHz}.  Low bandwidth reduces noise.  Also,
low bandwidth results in slow rise of signal upon charge arrival.  Combined with the large
$\tau_f$, which ensures long signal retention, the output of CSA can be multiplexed and sampled
slowly without going through an anti-alias filter or an in-pixel shaper.  The entire analog
channel is DC coupled.  The CSA output is also DC coupled to the discriminator.  DC coupling is
necessary to handle slowly varying signals.

The comparator compares the CSA output to a set threshold provided by a 4-bit DAC in each pixel.
All the 4-bit DACs in the entire array share a common offset that is settable externally.  The
finest step of each 4-bit DAC is tuned to be \SI{6}{mV} so that the entire dynamic range of the
DAC (\SI{90}{mV}) could cover the dispersion of the comparator input across the array.  When the
CSA output surpasses the threshold, a hit is generated and the Priority Logic Module is notified
by the \sym{Flag} signal.

There is a 5-bit SRAM in each pixel.  4 bits are for setting the value of the DAC.  The fifth bit
controls the digital \sym{EN}.  The SRAM can be written into when the pixel is selected during
analog multiplexing via \sym{ROW\_SEL} and \sym{COL\_SEL}.  SRAM was chosen over D Flip-Flop to
save the floor space.

\subsection{Analog Readout}

The analog output from the CSA in each pixel goes through two source-follower stages (Ms1 and Ms2
in Fig.~\ref{fig:SinglePixelStruct}) before entering the analog output buffer that drives the
signal off-chip for the entire array (lower dashed box in Fig.~\ref{fig:SinglePixelStruct}).  A
time-shared multiplexing scheme is used to readout the entire array.  A Scan Module controls the
row selection switch (\sym{ROW\_SEL}) and the column selection switch (\sym{COL\_SEL}) to select
one pixel at a time for its analog signal to pass through to the output buffer.  Each pixel has a
set of Ms1, Ms2, and the \sym{ROW\_SEL} switch.  Each column shares one \sym{COL\_SEL} switch.
The entire sensor has one output buffer and one pair of Ma1 and Ma2.  The first stage
source-follower Ms1 and its \SI{2}{\micro A} bias is constantly on for all the pixels while the
second stage source follower is only on for the single chosen pixel in multiplexing.  Since the
CSA has very limited drive strength, it is easily affected by the row and column switching.  The
addition of the first stage source-follower is necessary to isolate the switches from the CSA.

Switches operate at every rising edge of the clock fed into the Scan Module.  The behavior of the
analog output during pixel switching is shown in Fig.~\ref{fig:RawWavMuxing}.  The Scan Module
accepts a clock of frequency ranging from 0 to tens of \si{MHz}.  We could supply a controlled
number of cycles of clock, then completely stop the clock, in order to select a desired pixel to
be connected statically to the output buffer.  This feature was exploited during analog noise
measurements.

\subsection{Digital Readout}

When \sym{EN=1}, the Priority Logic Module in each pixel will respond to the output of the
comparator (\sym{Flag}), which becomes 1 when the CSA output exceeds the set threshold.  The
Priority Logic Module is a combinational logic that controls the reset (\sym{FB\_RST}) of the CSA
upon a hit and drives the hit information through the column readout structure.  Its behavior is
governed by Eq.~\ref{eq:1}
\begin{subequations}
  \label{eq:1}
  \begin{align}
    m &= \begin{cases}
      \sym{Flag}\wedge\sym{EN} & \text{if\ }\neg\sym{Reset}\vee\sym{AddrEN}=1\\
      0 & \text{else}
    \end{cases}\label{eq:1a}\\
    \sym{PFO} &= \sym{PFI}\vee m\label{eq:1b}\\
    \sym{AddrEN} &= \neg\sym{PFI}\wedge m\label{eq:1c}\\
    \sym{FB\_RST} &= \sym{Reset}\wedge\sym{AddrEN}\label{eq:1d}
  \end{align}
\end{subequations}
where $m$ is an intermediate variable.

Pixels are arranged and interconnected in columns.  For the $i$th pixel, its \sym{PFI}$_i$ is
connected to the previous ($(i-1)$th) pixel's \sym{PFO}$_{i-1}$, and its \sym{PFO}$_i$ is fed
into the next ($(i+1)$th) pixel's \sym{PFI}$_{i+1}$.  Pixels in the same column are daisy-chained
in this fashion.  Each pixel in the same column has a unique 7-bit address.  The address
controller in each pixel is connected to an address bus common to a column.  The address bus is
normally pulled up to all 1.  When \sym{AddrEN} is active in a pixel, said pixel pulls down the
address bus to its own unique address.  A common \sym{COL\_RST} is sent from the \sym{COL\_READ}
module to the \sym{Reset} port of every pixel in the column simultaneously
(Fig.~\ref{fig:SinglePixelStruct}).

The top-most pixel (0th) in a column has \sym{PFI}$_0=0$.  When there is no hit in any pixel, the
propagation of Eq.~(\ref{eq:1b}) dictates that every pixel in the column has
$\sym{PFI}=\sym{PFO}=0$.  When the $i$th pixel gets a hit, $m_i=1$ hence $\sym{PFO}_i=1$.  Due to
Eq.~(\ref{eq:1b}), all pixels below the $i$th pixel (denoted by $j$th, $j>i$) will have
$\sym{PFI}_j=\sym{PFO}_j=1$.  Due to Eq.~(\ref{eq:1c}), any pixel with $\sym{PFI}=1$ won't enable
\sym{AddrEN} even if it gets a hit.  This logic describes a priority chain of pixels: the pixel
with a hit that has the lowest $i$ (highest priority) could enable \sym{AddrEN}, and it disables
all the pixels lower in the chain from reacting to hits.  An associated consequence is that the
address bus is pulled down by only one pixel (the highest priority pixel with a hit) so that no
race condition rises on the address bus.

The \sym{COL\_READ} module monitors the address bus.  It reacts to the address change on the bus,
then reads the hit and resets the pixel.  A digital multiplexer (\sym{MUX} in
Fig.~\ref{fig:TMIImOverall} (right)) polls the status of each \sym{COL\_READ} module sequentially
once per clock cycle.  It assembles the hit pixel address and the hit time data, then ships them
off the sensor.

\begin{figure}[htbp]
  \centering
  \includegraphics[width=\linewidth]{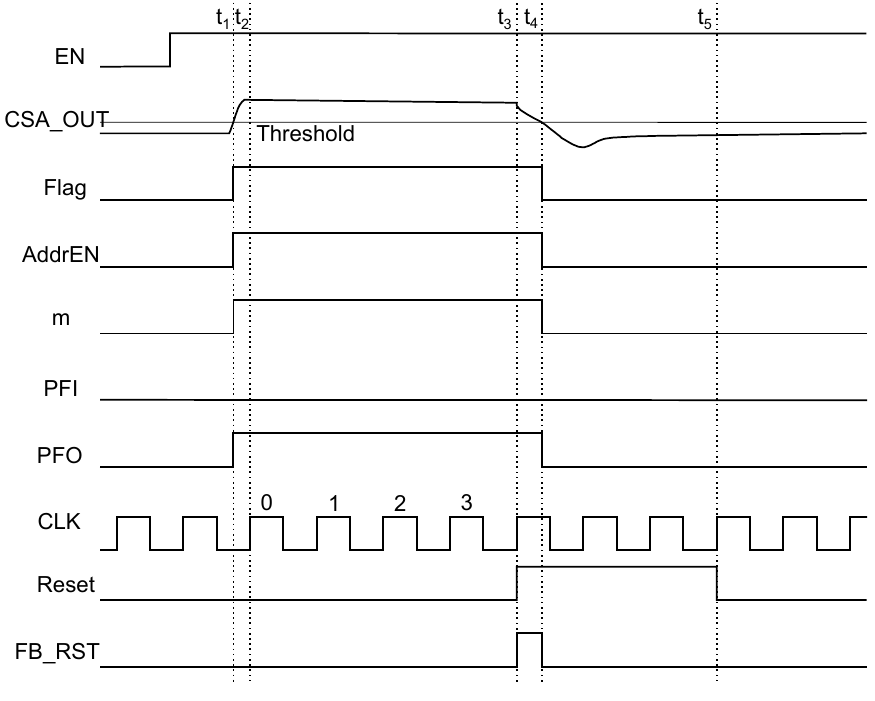}
  \caption{Timing diagram of relevant signals during a hit and its readout.  $t_1$: charges
    arrive and cause \sym{CSA\_OUT} to exceed the threshold; $t_2$: the \sym{COL\_READ} module
    senses a change on the address bus; $t_3$: the \sym{COL\_READ} module confirms the address
    change and sends a reset signal to the hit pixel; $t_4$: \sym{CSA\_OUT} falls below the
    threshold due to the reset; $t_5$: the \sym{COL\_READ} is polled and the reset is removed.}
  \label{fig:PriLogicSeq}
\end{figure}

A timing diagram of a hit-read-reset cycle is shown in Fig.~\ref{fig:PriLogicSeq}.  Charges
arrive at $t_1$, causing the CSA output to exceed the threshold, generating a hit
($\sym{Flag}=1$).  Activated by $\sym{EN}=1$, a single-pole-double-throw (\sym{SPDT}) switch
grounds the gate of Mf from its original bias \sym{FB\_VREF}, so the CSA maximally retains the
charge signal.  Since there is no higher-priority pixel ($\sym{PFI}=0$) and this pixel is enabled
($\sym{EN}=1$), $m$, \sym{PFO} and \sym{AddrEN} become 1 accordingly.  At this moment the address
bus is pulled to the address of this pixel as well.  At $t_2$ (rising edge of the clock in the
\sym{COL\_READ} module), the \sym{COL\_READ} module senses the address change; then, it waits for
4 clock cycles to make sure the address change is not a transient phenomenon.  At the end of the
waiting period, $t_3$, the \sym{COL\_READ} module registers the time from a system timer
(counter) and sends a reset signal to \sym{Reset}.  Although \sym{Reset} is sent to every pixel
in the column, due to Eq.~(\ref{eq:1d}), only the pixel that is pulling the address bus and is
being read out will respond to the reset.  The reset sets $\sym{FB\_RST}=1$, which turns on the
feedback transistor Mf to discharge $C_f$ so that the CSA output comes down towards the baseline.
At $t_4$, the CSA output falls below the threshold causing $\sym{Flag}=0$ hence \sym{FB\_RST} is
removed.  \sym{Reset} is removed when the \sym{COL\_READ} module is polled ($t_5$).  The time
between $t_4$ and $t_5$ is non-deterministic and can be as high as 72 clock cycles.

When multiple pixels in the same column are hit simultaneously, the logic reads out and resets
the hit pixels sequentially in an order following their priorities.  Eq.~(\ref{eq:1a}) ensures
that when a higher-priority hit pixel is reset, the next-priority hit pixel is not reset until
\sym{Reset} is toggled.  No hit is missed; however, the timing is only accurate for the pixel
with the highest priority.  The timing resolution is determined by the system timer (counter)
frequency.

It is worth noting that no digital clock is sent into the array.  Clock is only fed to the
\sym{COL\_READ} module at the bottom of each column.  The digital logic in the array is entirely
combinational and asynchronous.  This design minimizes the interference between digital logic and
the analog circuit.  Tests show that this readout implementation works correctly as designed.

Column-based readout structures have been realized in several ICs such as
\cite{Millaud1995,Peric2006,Hemperek2009}.  Unique addresses for pixels and the propagation of
priorities, sometimes called tokens, were also used.  The major difference in this design is that
a reset signal is required to be sent into the pixel being read out to bring the CSA output down
below the threshold.  We implemented the logic to handle this requirement correctly.

\section{Test results}

All the following tests were conducted at room temperature in ambient air.

\subsection{Analog noise measurement}

We applied a calibrated square wave on the guard ring.  The square wave has a peak-to-peak
amplitude of \SI{10}{mV}.  Since there is a coupling capacitance, \Cgt, between the \sym{Gring}
and the CSA input, at each transition edge of the square wave, an equivalent charge
$Q_i=\Cgt\times\SI{10}{mV}\approx\SI{343}{e^-}$ is injected into the CSA.  At a rising edge of
the square wave, positive equivalent charge is injected.  Negative equivalent charge is injected
at a falling edge.  The CSA responds to both polarities equally well; however, we focus on the
negative equivalent charge in this measurement.  The frequency of the square wave is chosen to be
low enough so that the CSA output has sufficient time to fall back to the baseline before the
next transition arrives.

\begin{figure}[htbp]
  \centering
  \includegraphics[width=\linewidth]{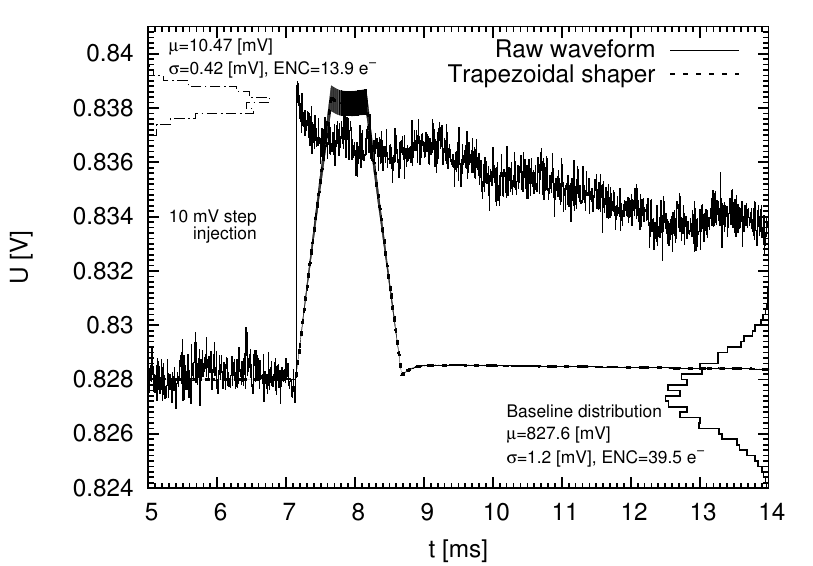}
  \caption{CSA noise measurement using pulse injection.  10 mV pulse steps (falling edges of a
    square wave) are repeatedly injected through \Cgt into the CSA.  The raw waveform is a
    snapshot of the CSA response to one pulse, sampled off the chip while the Scan Module stops
    at one pixel.  We applied a trapezoidal filter in software with \SI{1}{ms}
    Full-Width-Half-Maximum (FWHM) to each pulse and aligned the outputs in time to build up a
    histogram (dashed line and band).  The band around the shaper output shows the $\pm1\sigma$
    contour.  The distribution of the flat-top part is shown in the top-left corner.  The
    distribution of the baseline before the rising edge, from many pulses, is drawn in the
    bottom-right corner.  $\Vgsf=\SI{140}{mV}$, $\tau_f=\SI{7.6}{ms}$.}
  \label{fig:CSAPulse}
\end{figure}

We utilized the pixel selection feature of the Scan Module to stop at a pixel and digitize its
analog output continuously.  An example of the CSA output response to a negative charge injection
is shown in Fig.~\ref{fig:CSAPulse}.  Since the entire analog channel is DC coupled, the baseline
is at a value determined by \sym{CSA\_VREF} and level-shifts from source followers.  As expected,
the CSA output rises sharply upon the arrival of a pulse, then decays down towards the baseline
exponentially with a time constant determined by $\tau_f=R_f\cdot C_f$, which is in the range of
several milliseconds.

Traditionally the noise performance of a CSA is reported as the fluctuation of the pulse height
after a shaper.  Since we do not have a shaper in the sensor, we applied a digital trapezoidal
filter\cite{Jordanov1994} in software.  The trapezoidal filter automatically zeros the baseline
of its output.  We shifted the filter output to match the baseline of the CSA output for better
graphical comparison.  The flat-top part of the trapezoidal filter output is used to evaluate the
height of the pulse.  By collecting many pulses, we show that the pulse height has a mean value
$\mu=\SI{10.47}{mV}$ and a standard deviation $\sigma=\SI{0.42}{mV}$.  The charge conversion gain
is then evaluated as $Q_i/\mu=\SI{32.8}{e^-/mV}$ and the Equivalent Noise Charge (ENC) is
$Q_i\cdot\sigma/\mu=\SI{13.9}{e^-}$.

Although the amplitude of the square wave is calibrated, we acknowledge that \Cgt is not
independently calibrated and that its value \CgtVal is entirely from simulation albeit both the
IC design software and an FEA analysis gave consistent results.  Nevertheless, we report on the
noise measurement based on the simulation result of \Cgt and note it as the sole source of
uncertainty in the noise value.  The reported noise includes noise from the entire data
acquisition chain; therefore, it is an upper limit of the true noise in the sensor.

Using the same data, by collecting many baseline samples before the rise in the CSA output, we
obtain a standard deviation of \SI{1.2}{mV} of the baseline, which is equivalent to
\SI{39.5}{e^-} using the above computed charge conversion gain.  The baseline fluctuation
determines the minimum threshold of the discriminator.  In this case, the finest adjustable step
of the threshold is \SI{6}{mV}, determined by the in-pixel DAC, which is 5 times the standard
deviation of the baseline.  Therefore we estimate the minimum threshold to be \SI{200}{e^-}.

\subsection{Alpha induced charge tracks in ambient air}

\begin{figure}[htbp]
  \centering
  \includegraphics[width=0.6\linewidth]{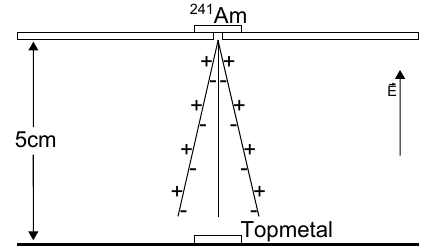}
  \caption{Experimental setup.  An $^{241}$Am alpha source is placed on top of a metal plate
    biased at \SI{-1}{kV} and \SI{5}{cm} above a \TMIIm sensor.  The alpha tracks are coarsely
    collimated by a small through-hole in the plate.}
  \label{fig:Am241Setup}
\end{figure}

We installed a thin metal plate, biased at \SI{-1}{kV} relative to the sensor ground, \SI{5}{cm}
above and parallel to the top surface of a \TMIIm sensor.  The entire setup is in ambient air
(Fig.~\ref{fig:Am241Setup}).  A nearly uniform electric field of \SI{200}{V/cm} is generated
between the plate and the sensor.  A $\sim\SI{0.5}{mm}$ through-hole in the plate is aligned with
the center of the sensor.  A spectroscopic $^{241}$Am alpha source with a thin window is placed
on the plate above the through-hole so that alpha particles could travel through the hole towards
the sensor and ionize air along the tracks.  The through-hole coarsely collimates the alpha
particles to a general downward-going direction towards the sensor; however, it allows some
divergence so that alphas have a probability to travel sideways.  The $^{241}$Am alpha source
emits \SI{5.45}{MeV} alphas that range out at about \SI{4}{cm} in air; therefore, no alpha
particle directly hits the sensor.  The ionization charges, which are believed to be mostly ions
in air, drift slowly in the electric field at a speed of several \si{mm/ms} \cite{Charpak2008}.

\begin{figure}[htbp]
  \centering
  \includegraphics[width=\linewidth]{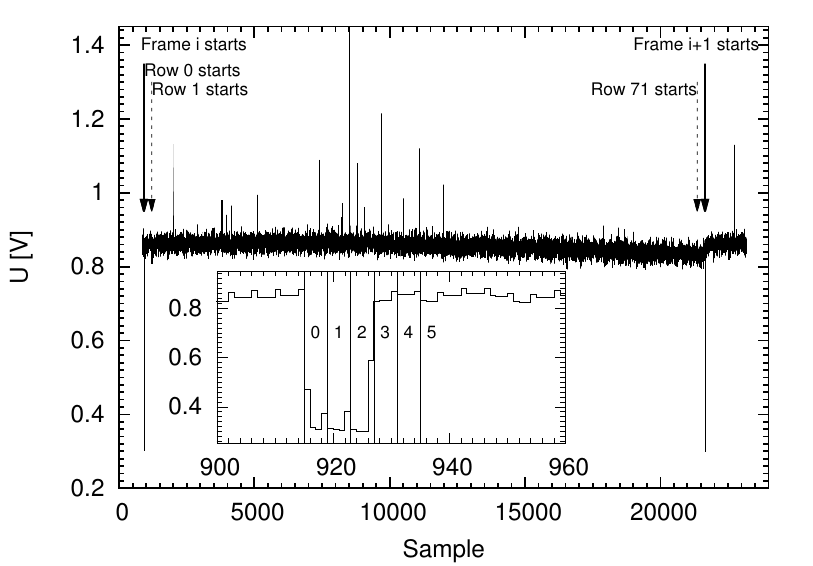}
  \caption{Analog output of the array during pixel multiplexing.  One full frame is shown.  Inset
    shows a zoomed-in view of the beginning of a frame.  The Scan Module controls the array
    scanning using the row-major order.  Each pixel is sampled 4 times and the average of the
    middle two samples is used as the voltage of the pixel in software.  Pixels in row 0, column
    0,1,2 are marker pixels tied to a fixed low voltage to facilitate the identification of the
    start of a new frame in analog waveform.}
  \label{fig:RawWavMuxing}
\end{figure}

We operated the \TMIIm sensor at an identical setting as in the noise measurement,
$\Vgsf=\SI{140}{mV}$.  Digital modules were entirely turned off by setting $\sym{EN}=0$ in each
pixel.  A \SI{7.8125}{MHz} clock was supplied to the Scan Module to drive the analog
multiplexing.  Under such clock frequency, each pixel occupies \SI{128}{ns} in the analog output
of the array, and it takes \SI{0.6636}{ms} to scan all of the $72\times72$ pixels once (one
frame).  Effectively each pixel is sampled once every \SI{0.6636}{ms}, and each sampling lasts
\SI{128}{ns}.  We used an external digitizer working at \SI{31.25}{Msps} to record the
multiplexed analog output.  The Scan Module clock is derived from the digitizer clock at $4:1$
ratio; hence, the two clocks are synchronized, and so each pixel is digitized exactly 4 times.
We took the average of 2 samples in the middle as the digitization value for the pixel discarding
the leading and trailing samples to eliminate the transition period at pixel switching.  A data
sample illustrating this sampling scheme is shown in Fig.~\ref{fig:RawWavMuxing}.

\begin{figure}[htbp]
  \centering
  \includegraphics[width=\linewidth]{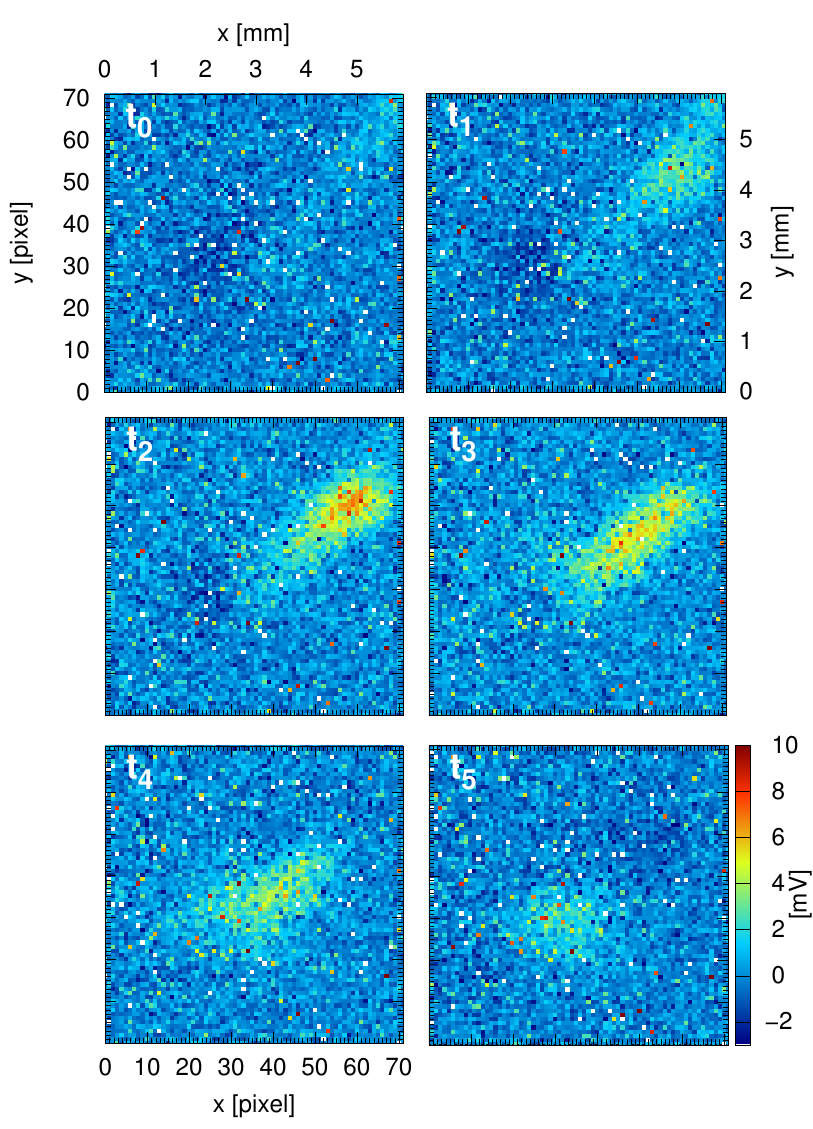}
  \caption{Time slices of a charge track generated by an alpha particle from an $^{241}$Am source
    ionizing ambient air.  Time progresses from $t_0$ to $t_5$ at equal interval.  The time
    between consecutive images is about \SI{3.3}{ms}.}
  \label{fig:AlphaImg}
\end{figure}

After de-multiplexing in software, we obtain a continuous waveform for each pixel sampled at a
rate of $1/(\SI{0.6636}{ms})\approx\SI{1.5}{kHz}$.  There are $72\times72=5184$ independent
waveforms for the entire array.  Since the signal retention time of CSA is long (milliseconds),
such low effective sampling rate is sufficient to capture charge signals.  We applied a software
trapezoidal filter to each waveform independently.  For a given time (sample), the height of
filter outputs from every pixel form an image.  Alpha particle induced charge tracks are
identified in these time dependent images.  A set of images of a single track is shown in
Fig.~\ref{fig:AlphaImg}.

The setup constrains alpha tracks to be mostly perpendicular to the sensor surface with some
inclination angle.  When the leading part of the charge track arrives at the sensor, other parts
are still away from the sensor.  Due to the slow-moving nature of charges drifting in air, even
at a sampling rate as slow as \SI{1.5}{kHz}, the charge density variation due to the charge
arrival time difference is visible in the data.

\section{Summary}\label{sec:summary}

We have demonstrated the implementation of a highly pixelated sensor for direct charge collection
and imaging using a standard \SI{0.35}{\micro m} CMOS technology.  The combination of direct
charge collection, low noise, and long signal retention makes the sensor appealing in several
low background and low event-rate density applications involving slow-drifting ions without
gaseous avalanche gain.  A column based digital readout structure also allows hits to be
registered efficiently.

To improve beyond \TMIIm, besides increasing the pixel density, we can further reduce the noise
of the CSA.  We will explore these options in future series of \TM sensors.

\section*{Acknowledgments}

This work is supported, in part, by the Thousand Talents Program at Central China Normal
University and by the National Natural Science Foundation of China under Grant No.~11375073.  We
also acknowledge the support from LBNL for hosting the physical measurements of the sensor.  We
would like to thank Christine Hu-Guo and Nu Xu for fruitful discussions.



\bibliographystyle{elsarticle-num-names}
\bibliography{refs}

\end{document}

%% file: defs.tex
\usepackage{xspace}

\newcommand{\TM}{\emph{Topmetal}\xspace}
\newcommand{\TMI}{\mbox{\emph{Topmetal-I}}\xspace}

\newcommand{\TMIIm}{\mbox{\emph{Topmetal-II\raise0.5ex\hbox{-}}}\xspace}

\newcommand{\znbb}{\ensuremath{0\nu\beta\beta}\xspace}
\newcommand{\Cgt}{\ensuremath{C_\text{GT}}\xspace}
\newcommand{\CgtVal}{\SI{5.5}{fF}\xspace}
\newcommand{\CinVal}{\SI{23}{fF}\xspace}
\newcommand{\Vgsf}{\ensuremath{V_\text{gs}^f}\xspace}

\DeclareSIUnit\keV{keV}
\DeclareSIQualifier\ee{ee}
\DeclareSIQualifier\nr{nr}
\DeclareSIUnit\pe{p.e.}

\newcommand{\sym}[1]{\texttt{#1}}
